\documentclass[prl,amsmath,twocolumn,showpacs]{revtex4}
\usepackage{bm}
\usepackage{amssymb}
\usepackage{graphicx}
\usepackage{amsmath,graphicx}
\begin{document}
\def\la{{\langle}}
\def\ra{{\rangle}}
\def\Epsilon{{\mathcal{E}}}
\title{Qubit residence time measurements with a Bose-Einstein condensate}
%
%
\author{D.Sokolovski}
\address{School of Mathematics and Physics,
  Queen's University of Belfast,
  Belfast, BT7 1NN, United Kingdom}
\begin{abstract}
We show that an electrostatic qubit located near a Bose-Einstein condensate
trapped in a symmetric double-well potential can be used to measure the duration
the qubit has spent in one of its quantum states. The strong, medium and weak 
measurement regimes are analysed and a new type of Zeno effect is discussed.
The analogy between the residence and the traversal (tunnelling) times is highlighted.
\end{abstract}

%
%
\maketitle
%
With the recent progress in quantum information technology there often 
arises a necessity to measure and control the state of a two-level
quantum system (qubit).
This can be achieved by constructing hybrid devices in which 
a microscopic irreversible current between two reservoirs is 
effectively controlled by the qubit's quantum state.
Such a device can be realised, for example, by placing an electrostatic qubit 
close to a point contact (PC) \cite{PC1}-\cite{PC3} or an non-interacting Bose-Einstein condensate trapped in a symmetric optical dipole trap \cite{BEC}. 
The two systems have been shown to affect the observed qubit differently: 
whereas a PC converts any qubit's initial state $|i\ra$ into a statistical
mixture exponentially in time \cite{GURV}, decoherence of a qubit coupled to a
BEC is much slower ($\sim 1/\sqrt{t}$) and strongly dependent on the choice 
of $|i\ra$ \cite{BEC}. 
While in a  PC set up one measures  the current across the
contact, a BEC device is best suited for observing the number of atoms
which have tunnelled into a previously empty reservoir after a time T.
The purpose of this Letter is to demonstrate that a symmetric BEC device 
whose Rabi oscillations are effectively blocked by the presence of 
the electron in the first qubit's dot, 
performs a quantum measurement of the qubit's residence time, i.e., 
the net duration the electron has spent in the second dot between $t=0$ and $t=T$.  
We will show that conceptually the question of residence time is closely related to the
traversal (tunnelling) time problem
 still actively debated
in the literature (see, for example, \cite{TRAV1}-\cite{TRAV}) 
In both cases the time in question is the duration a system spends
in a specified sub-space of its Hilbert space, the sub-barrier region
or the state in one of the quantum dots.
Both quantities relate to  
to the total duration of the system's motion, rather than to a single 
instant, and are conveniently represented by a Feynman functional. 
One can extend von Neumann's measurement theory to such functionals \cite{SR}, 
but, as far as we know, the BEC device proposed in this Letter offers the
first practical realisation of such a measurement. 
\newline
The role of a BEC as a measurement tool is best illustrated 
by considering first
somewhat simpler case of a condensate coupled to a two-level fluctuator, 
i.e. a classical bistable system switching randomly between two  
positions so that its path $q(t)$ is a random function taking values
of either $0$ or $1$ \cite{FLUC}. Assuming that the tunnelling rate of the BEC atoms
is enhanced (the barrier is lowered) when $q=1$, we write the Hamiltonian as
\begin{eqnarray} \label{0.1}
\hat{H}_{BEC}(t)= [\Omega + q(t) \delta \Omega)](c_L^+c_R+c_R^+c_L), \quad \delta \Omega > 0
\end{eqnarray}
The condensate consists of $N$ atoms initially (at $t=0$)  located in the left well.
After a time $T$ we wish to count the number of atoms in the right well, $n$,
in order to obtain information about the noise $q(t)$ for 
$0\le t\le T $.
 The probability amplitude for $n$ atoms to tunnel into the right well 
is a functional on the fluctuator's path $q$
given by
\begin{eqnarray} \label{0.1a}
G_{n\leftarrow 0}[q(\centerdot)]= \la n|\exp[-i(\Omega T + \delta \Omega \tau)(c_L^+c_R+c_R^+c_L)]|0\ra
\end{eqnarray}
where $|n\ra$ denotes the BEC state with $n$ atoms in the right well, and $\tau$ 
is the duration the fluctuator has spent in the state $q=1$, explicitly given
by an expression similar to the traversal time functional 
$\tau_{ab}[x(\centerdot)]$ 
of Ref. \cite{TRAV1} 
($\delta_{ij}$ is the Kroneker delta)
\begin{eqnarray} \label{0.3}
\tau_1[q(\centerdot)]\equiv \int_0^T \delta_{q(t)1}dt.
\end{eqnarray}
For simplicity we will assume that no tunnelling occurs for $q=0$,
i.e. $\Omega=0$, and the BEC consists of a large number of identical 
non-interacting  atoms whose
Rabi period $2\pi/\delta \Omega$ is large compared to the observation time $T$,
\begin{eqnarray} \label{0.4}
N\rightarrow \infty, \quad \delta \Omega \equiv \alpha /N^{1/2} \rightarrow 0, \quad
 \delta \Omega T \rightarrow 0.
\end{eqnarray} 
Condition (\ref{0.4}) ensures that if the barrier is permanently lowered,
($q(t) \equiv 1$), there is an irreversible macroscopic current into the right reservoir, with the number of tunnelled atoms increasing as $\alpha^2 t^2$.
The spectrum 
of the operator in the exponent of Eq.(\ref{0.1a}) consists of equidistant 
levels, $\epsilon _{n} =(2n-N)\delta \Omega$, $n=0,1..N$. Expanding the exponential in Eq.(\ref{0.1a}) in the basis of the corresponding eigenstates and using the 
Sterling formula for the factorials yields
\begin{eqnarray} \label{0.5}
G_{n\leftarrow 0}(\tau) \approx \alpha^n\tau^n \exp(-\alpha^2\tau^2/2)/(n!)^{1/2} \\
\nonumber
 \rightarrow _{n>>1}
(2\pi n)^{-1/4} exp[-\alpha^2(\tau-\tau_n)^2].
\end{eqnarray}
where      
\begin{eqnarray} \label{0.6}
\tau_n\equiv n^{1/2}/\alpha
\end{eqnarray}
is the time after which on average $n$ atoms escape to the 
right well.
%
If the fluctuator's paths are distributed with a functional density $W[q(t)]$,
the probability to find $n$ atoms in the right well, $ 
|G_{n\leftarrow 0}(\tau)|^2$, must be averaged further, and we obtain
\begin{eqnarray} \label{0.7}
P_{n\leftarrow 0}(T)= \int_0^T |G_{n\leftarrow 0}(\tau)|^2 W(\tau,T) d\tau  
\end{eqnarray}
where the restricted path sum
\begin{eqnarray} \label{0.7a}
W(\tau,T) \equiv \sum_{paths} \delta(\tau-\tau_1[q(\cdot)])W[q(t)]
\end{eqnarray}
is the fluctuator residence time probability distribution. 
Thus, finding at $t=T$ exactly $n$ tunnelled atoms allows us 
to conclude that the fluctuator has kept the barrier open for a duration 
$\tau_n - 1/\sqrt{2}\alpha < \tau < \tau_n + 1/\sqrt{2}\alpha$, i.e. that we have measured
its residence time to an accuracy $\Delta \tau = 1/\sqrt{2}\alpha$.  
\newline 
Next we replace the fluctuator with a qubit  placed near the BEC dipole trap  in such a way
that the BEC tunnelling rate is enhanced whenever qubit's  electron is located in the state $|1\ra$.
The Hamiltonian of the system
can be written as $\hat{H}=\hat{H}_q + \hat{H}_{BEC}$, where
\begin{eqnarray} \label{0.8}
\hat{H}_q(\epsilon)=\epsilon a_1^+ a_1+ \omega(a_1^+a_2+a_2^+a_1)\quad\quad\quad\quad\quad\quad \quad\\
\nonumber
\hat{H}_{BEC}=(\Omega + a_1^+a_1 \delta \Omega)(c_L^+c_R+c_R^+c_L),\quad \quad \delta \Omega > 0 
\end{eqnarray}
and $a_{1(2)}^+$ are the creation operators for the qubit's electron electron in the first
(tunnelling enhanced) and the second (tunnelling suppressed) quantum dot, respectively.
In the following we will put the qubit's Rabi frequency to unity, 
$\omega=1$, and re-scale other time and energy parameters accordingly.
Like a two-state fluctuator, a qubit can alternate between the two states, $|1\ra$ and $|2\ra$,
with the important difference that its trajectory $q(t)$ taking the values $1$ or $2$ is a virtual (Feynman)
path.  To such a path one can assign a probability amplitude $\Phi[q(t)]$ but not, as above, a probability weight $W[q(t)]$. 
We must, therefore, evaluate the number of tunnelled atoms at $t=T$
without being able to predict, even with a probability,
whether the barrier was up or down at any previous time $0\le t<T$ \cite{G2}. 
For a qubiit starting its motion (pre-selected)  in the state $|i\ra$ and
then at $t=T$ observed (post-selected) in a final state $|f\ra$, this probability amplitude is given by  $\Phi^{f\leftarrow i}[q]=\la f|q(T)\ra (-i\omega)^j\exp(-i\epsilon \tau) \la q(0)|i\ra$,
where $j$ is the number of times the path crosses from one state to another.
Following Feynman and Vernon \cite{VF} we can obtain the probability amplitude $A^{f\leftarrow i}_{n\leftarrow 0}(T)$  for finding  $n$ atoms in the right  well given the initial and final states of the qubit
by multiplying the amplitude in Eq.(\ref{0.5}) by $\Phi^{f\leftarrow i}[q]$ and summing over all qubit's paths. Assuming, as above, $\Omega =0 $ and recalling that $G_{n\leftarrow 0}[q(t)]$ only depends on the path's residence time (\ref{0.3}), we write 
\begin{eqnarray} \label{0.9}
A^{f\leftarrow i}_{n\leftarrow 0}(T)= \int_0^T G_{n\leftarrow 0}(\tau) \Phi^{f\leftarrow i}(\tau,T) d\tau  
\end{eqnarray}
where the restricted path sum (c.f. Refs. \cite{TRAV1})
\begin{eqnarray} \label{0.9a}
 \Phi^{f\leftarrow i}(\tau,T) \equiv \sum_{paths} \delta(\tau-\tau_1[q(\cdot)])\Phi^{f\leftarrow i}[q(t)]
\end{eqnarray}
is the qubit's residence time probability amplitude distribution. Thus, the  quantum analogue of Eq.(\ref{0.7}) is
\begin{eqnarray} \label{0.10}
P_{n\leftarrow 0}^{f\leftarrow i}(T)= |\int_0^T G_{n\leftarrow 0}(\tau) \Phi^{f\leftarrow i}(\tau,T) d\tau|^2. 
\end{eqnarray}
From Eq.(\ref{0.5}) it is readily seen that  the probability $P_{n\leftarrow 0}^{f\leftarrow i}$
results from the interference between the paths with $\tau_n-1/\alpha \lesssim  \tau \lesssim
\tau_n+1/\alpha $,
so that by determining $n$
we perform a measurement of the qubit's residence time \cite{FOOT1} to a {\it quantum} accuracy 
$\Delta^q \tau  \equiv 1/\alpha$ \cite{FOOT2}.
Finally,
if the maximum number of atoms which can tunnel over the time $T$ is large,
$N_{max}\approx \alpha^2T^2 >>1$, we can introduce probability 
density $w^{f\leftarrow i}(\tau,T)$ for the measured values of $\tau$,
$w^{f\leftarrow i}(\tau,T)\equiv P_{n\leftarrow 0}^{f\leftarrow i}(T)(d\tau_n/dn)^{-1}=2\alpha \sqrt{n} P_{n\leftarrow 0}^{f\leftarrow i}$.
Explicitly we have
%
\begin{eqnarray} \label{0.12}
w^{f\leftarrow i}(\tau,T) \approx (2/\pi)^{1/2} \alpha\quad\quad\quad\quad\quad\quad\quad\quad\quad\quad \\
\nonumber
\times |\int_0^T \exp[-\alpha^2(\tau-\tau')^2]\Phi^{f\leftarrow i}(\tau',T)d\tau'|^2.
\end{eqnarray}
\newline
The measurement statistics are determined by the distribution  (\ref{0.9a}),
some of whose properties have been discussed in  \cite{SRES}.
In particlular, it follows from Eq.(\ref{0.9a}) that 
\begin{eqnarray} \label{0.10a}
\Phi^{f\leftarrow i}(\tau,T)=\quad\quad\quad\quad\quad\quad\quad\quad\quad\quad\quad\quad\quad\quad\\
\nonumber
(2\pi)^{-1}\exp(-i\epsilon \tau)\int \exp(i\lambda \tau)
\la f|\hat{U}(T,\lambda)|i\ra d\lambda
\end{eqnarray}
where $\hat{U}(T,\lambda)$ is the evolution operator for an asymmetric qubit with 
the Hamiltonial $\hat{H}_q(\lambda)\equiv  \lambda a_1^+a_1+(a_1^+a_2+a_2^+a_1)$,
whose matrix elements, $U_{kk'}\equiv \la k|\hat{U}(T,\lambda)|k'\ra$ are given by 
\begin{eqnarray} \label{0.11}
\nonumber
U_{11} = [\cos(\mathcal{E}T/2)
-i \lambda \Epsilon ^{-1} \sin(\Epsilon T/2)]
\exp(-i\lambda T/2)\\
\nonumber 
\equiv \exp(-i\lambda T)+u_{11}(\lambda)\\
\nonumber
U_{22} = [\cos(\mathcal{E}T/2)
+i\lambda \mathcal{E}^{-1} \sin(\Epsilon T/2)]
\exp(-i\lambda T/2)\\
\equiv 1+u_{22}(\lambda)\quad\quad\quad\quad \\
\nonumber
U_{12}=
-2i \Epsilon^{-1} \sin(\Epsilon T/2)\exp(-i\lambda T/2)=U_{21}
\quad\quad
\end{eqnarray}
where $\Epsilon(\lambda) \equiv (\lambda^2+4)^{1/2}$ and $u_{11(22)}(\lambda)\rightarrow 0$ for $|\lambda|\rightarrow \infty$.
Inserting Eqs. (\ref{0.11}) into Eq. (\ref{0.10a}) shows that
\begin{eqnarray} \label{0.15a}
\Phi^{1\leftarrow 1}(\tau,T)=\delta(\tau-T)+\phi^{1\leftarrow 1}(\tau,T)
=\Phi^{2\leftarrow 2}(T-\tau,T) \nonumber\\
\Phi^{2\leftarrow 1}(\tau,T)=\phi^{2\leftarrow 1}(\tau,T)=\Phi^{1\leftarrow 2}(\tau,T) \quad\quad\quad
\end{eqnarray}
where $\delta (z)$ is the Dirac delta-function and  $\phi^{f\leftarrow i}$ are smooth functions of $\tau$.
For $T>>1$ $\phi^{f\leftarrow i}$  can be evaluated by the stationary
phase method \cite{STAT}. Considering for simplicity a symmetric qubit, $\epsilon=0$,
and introducing a new variable $\xi \equiv \tau/T-1/2$, $0\le \xi \le 1$,
we obtain the large-time semiclassical asymptotes valid 
for $0< \tau < T$,  
\begin{eqnarray} \label{0.12a}
\phi^{1\leftarrow 1}(\tau,T) \approx
(2/\pi T)^{1/2}\times \quad\quad\quad\quad\quad\quad \quad\quad\quad\quad\\
\nonumber 
(1+2\xi)^{1/4}(1-2\xi)^{-3/4}\cos[(1-4\xi^2)^{1/2}T+\pi/4]\quad
\end{eqnarray}
\begin{eqnarray} \label{0.12c}
\phi^{1\leftarrow 2}(\tau,T)
\approx\quad\quad\quad\quad\quad\quad\quad\quad\quad\quad
\quad\quad\quad\quad\quad\quad\\
\nonumber
-i(2/\pi T)^{1/2}(1-4\xi^2)^{1/2}\sin[(1-4\xi^2)^{1/2}T+\pi/4].
\end{eqnarray}
The oscillatory distributions $\Phi^{1\leftarrow 1}(\tau,T)$ and $\Phi^{1\leftarrow 2}(\tau,T)$ are shown in Fig.1. 
It is readily seen that after many Rabi periods of the qubit, 
$T>>1$, $\Phi^{1\leftarrow 1}$ 
develops
 a stationary region of the width
$\sim T^{1/2}$ centred at $\tau=T/2$, which suggests that on average the qubit shares its time equally between the states $|1\ra$ and $|2\ra$.
At the same time the singular term $\delta (\tau -T)$ appears to imply that the qubit has never left the state $|1\ra$. 
\begin{figure}[h]
\includegraphics[angle=0,width=8cm]{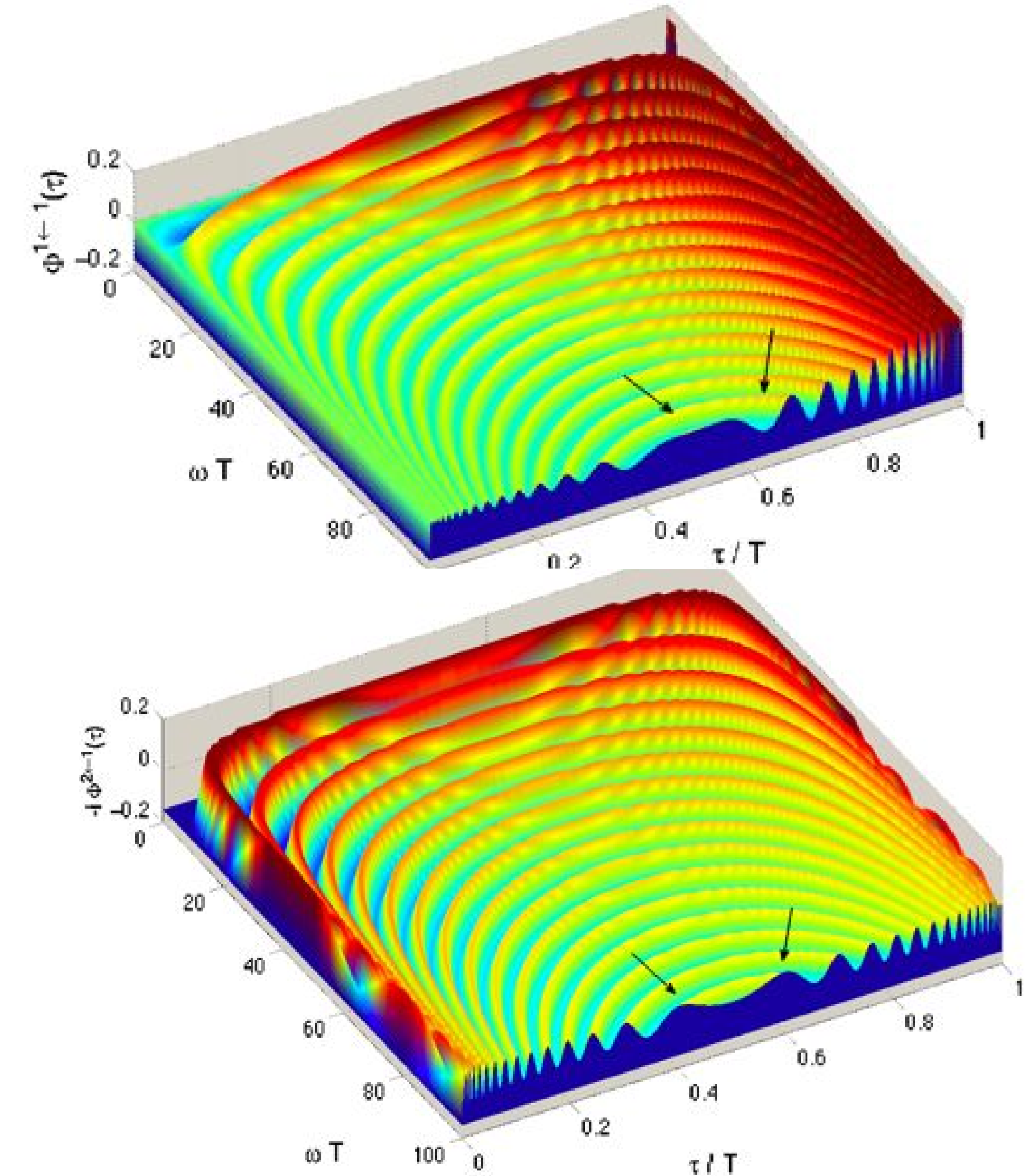}
\caption { Residence time amplitude distributions
$\Phi^{1\leftarrow 1}(\tau)$  and $\Phi^{2\leftarrow 1}(\tau)$ vs. $T$ and $\tau$ for
$0.02 \le \tau /T \le 0.98$. Arrows indicate the stationary phase region.}
\label{fig:FIG1}
\end{figure}
There is, however, no contradiction and next we will show that 
the two conflicting scenarios correspond to two different accuracies
of the BEC meter and, therefore, are never observed at the same time.

Indeed, for a medium accuracy, $T^{1/2} < \Delta^q \tau < T$, the main
contribution to integral (\ref{0.12}) comes from the stationary region in Fig.1.
and we have 
\begin{eqnarray} \label{0.16}
w^{1\leftarrow 1}(\tau,T) = w^{2\leftarrow 2}(\tau,T) =
\quad\quad\quad\quad\quad \quad\quad\quad\quad\quad\\
\nonumber
(2/\pi)^{1/2}\alpha \cos^2(T)\exp[-2\alpha^2(\tau-T/2)^2]\\
\nonumber
w^{1\leftarrow 2}(\tau,T) = w^{2\leftarrow 1}(\tau,T) =\quad\quad\quad\quad\quad\quad\quad\quad\quad\quad\\
\nonumber 
(2/\pi)^{1/2}\alpha \sin^2(T)\exp[-2\alpha^2(\tau-T/2)^2].
\end{eqnarray}
The Gaussian distributions (\ref{0.16}), shown in Fig.2a
for $\omega T=100$ and $\Delta^q\tau/T=0.1$
 by dashed lines,  are consistent with the qubit 
spending in the state $|1\ra$ roughly half of the total time $T$.
Note that here the contribution from the $\delta (\tau-T)$ term is cancelled 
by the oscillations of the regular part of $\Phi^{1\leftarrow 1}(\tau,T)$
near $\tau\approx T$.
To model an actual measurement and check the accuracy of Eqs.(\ref{0.12})
 and (\ref{0.16}) we have divided
the time interval $[0,T]$ into $N_{bin}=100$ equal subintervals $\delta t =T/N_{bin}$
, summed the 
probabilities $P_{n\leftarrow 0}^{f\leftarrow i}$ in Eq.(\ref{0.10}) within each interval and divided the
sum by $\delta t$. The results of this binning procedure are shown in Fig.1a by the solid lines.
\begin{figure}[h]
\includegraphics[angle=0,width=7cm]{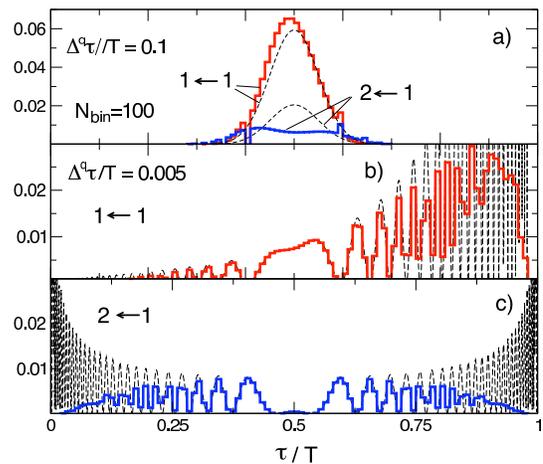}
\caption {a) Distributions $w^{f\leftarrow i} (\tau,T)$ for
$\omega T=100$ and
$\Delta \tau^q/T=0.1$: Eq.(\ref{0.16}) (dashed) and the binning procedure
with $N_{bin}=100$ (solid);
b and c) Same as (a) but
for $\Delta \tau^q/T=0.005$.}
\label{fig:FIG2}
\end{figure}
As we increase the coupling strength $\alpha$, the integral (\ref{0.12}) will 
still vanish wherever oscillations of $\Phi$ are fast compared to $\Delta^q \tau=1/\alpha$.
 Where $\Delta^q \tau$ is small compared to the oscillation's period
we obtain
\begin{eqnarray} \label{0.17}
w^{f\leftarrow i}(\tau,T) \approx (2\pi)^{1/2} \alpha^{-1}
|\Phi^{f\leftarrow i}(\tau,T)|^2,\quad 0<\tau<T.
\end{eqnarray}
Thus, as the accuracy improves, the measurement will resolve  
the pattern of $|\Phi^{f\leftarrow i}(T,\tau)|^2$ in ever greater 
detail. We also note that the probability densities 
in Eq.(\ref{0.17})
decrease as $\alpha^{-1}$ as
 interaction with BEC suppresses qubit's transitions between the states $|1\ra$ 
and $|2\ra$. 
The approximation (\ref{0.17}) and the results of a  binning procedure
with $N_{bin}=100$
are shown in Figs.2b and 2c for $\Delta^q \tau/T=0.005$ by the dashed and the solid lines, respectively. 
\newline
In the high accuracy limit $\alpha \rightarrow 0$ the probability is conserved owing to 
the $\delta(\tau-T)$ and $\delta(\tau)$ terms present, as seen from Eqs.(\ref{0.15a}),
 in $\Phi^{1\leftarrow 1}(T,\tau)$ and $\Phi^{2\leftarrow 2}(T,\tau)$,
respectively. Inserting them into Eq.(\ref{0.12}) shows that while
 $w^{1\leftarrow 2}(\tau,T)$ and  $w^{2\leftarrow 1}(\tau,T)$ vanish,
$w^{1\leftarrow 1}(\tau,T)$ and  $w^{2\leftarrow 2}(\tau,T)$ 
become
%
\begin{eqnarray} \label{0.18}
lim_{\alpha\rightarrow \infty} w^{1\leftarrow 1}(\tau,T)=\quad\quad\quad\quad\quad\quad\quad\quad\quad\quad\\
\nonumber (2/\pi)^{1/2}\alpha \exp[-2\alpha^2(\tau-T)^2]
\approx \delta(\tau-T)\\
\nonumber
lim_{\alpha\rightarrow \infty} w^{2\leftarrow 2}(\tau,T) 
\approx \delta(\tau)\quad\quad\quad\quad\quad\quad\quad\quad
\end{eqnarray}
indicating that the qubit is trapped in its initial state. 
Note that 
 no atoms will tunnel for a qubit starting in the second state, $|i\ra = |2\ra$,
whereas for $|i\ra = |1\ra$ one obtains
a narrow Poisson distribution
$
lim_{\alpha \rightarrow \infty}P_n^{1\leftarrow 1} = 
(\alpha T)^{2n}\exp(-\alpha^2T^2)/n!.
$
This variant of Zeno effect, which arises not from frequent observations
of the measured system \cite{ZENO},
but from one accurate evaluation the functional (\ref{0.3}) over a finite period of time, 
should be 
common to measurements of other quantities which are non-local in time.
\newline
In order to avoid the back action of the BEC on the qubit's evolution and
find the 'unperturbed' residence time one may be tempted to decrease the
coupling by putting $\alpha \rightarrow 0$. Again, it is instructive to analyse first 
the case of a classical fluctuator. In this weak coupling limit Eq.(\ref{0.5}) yields
$G_{n\leftarrow 0}(\tau)\approx \alpha^n \tau^n/\sqrt{n!}$ and from Eq.(\ref{0.7}) we obtain
$P_{n\leftarrow 0}(T)\approx \alpha ^{2n}
\la\tau ^{2n}\ra/n!$ where 
$\la \tau ^{2n}\ra \equiv \int_0^T\tau^{2n}W(\tau)d\tau$ 
is the $n$-th even moment of 
the probability distribution $W(\tau) \ge 0$. Thus from the ratio 
$P_{1\leftarrow 0}(T)/P_{0\leftarrow 0}(T) \approx \alpha^2\la\tau^2\ra$
we can determine $\la\tau^2\ra$ and, should the dispersion  be small,
the mean residence time $\la\tau\ra\approx \sqrt{\la\tau^2\ra}$.
\newline
Similarly, for a qubit from Eqs.(\ref{0.5}), (\ref{0.10}) and (\ref{0.10}) in the limit
$\alpha \rightarrow 0$ we find
 $P_n^{f\leftarrow i}/P_0^{f\leftarrow i}\approx \alpha^{2n} |\bar{\tau^n}|^2/n!$ with
$
 \overline{\tau^n} \equiv \int \tau^n \Phi^{f\leftarrow i}(\tau,T)d\tau/ \int  \Phi^{f\leftarrow i}(\tau,T)d\tau= 
(-i)^n\partial_\lambda 
log\la f|\hat{U}(T,\lambda)|i\ra|_{\lambda=0}.
$
 In particular, we have
   \begin{eqnarray} \label{0.15}
P_{1\leftarrow 0}^{1\leftarrow 1}/P_{0\leftarrow0}^{1\leftarrow 1} \approx \alpha^2 |\overline{\tau}|^2 =  \alpha ^2|T/2+\tan(T)/2|^2.
\end{eqnarray}
where $\overline {\tau}$ is the weak value of the residence time analogous  
to the Larmor tunnelling time first introduced to quantum scattering by Baz' \cite{LARM}.
%
It diverges
whenever Rabi oscillations put the unperturbed qubit into the state $|2\ra$, $T=(k+1/2)\pi$, 
$k=0,1...$, may exceed the total duration of motion $T$,
and cannot be interpreted as a valid residence time.
This problem is common to all weak measurements introduced in \cite{WEAK}, whose accuracy is so poor
that they do not destroy coherence between different values of the measured quantity \cite{SWEAK}. 
The weak residence time $\bar{\tau}$ in Eq.(\ref{0.15})
is the first moment of
an alternating amplitude distribution
$\Phi^{1\leftarrow 1}(T,\tau)$ and as such is not directly linked to the physical  values
$0\le \tau \le T$ \cite{SWEAK}.
\newline
In summary, we have shown that a hybrid device consisting of  an electrostatic
qubit coupled to a BEC trapped in a symmetric double-well potential can be
used to perform the qubit's residence time measurements. Depending on the strength of the coupling the measurement can be 'weak' or strong.
An accurate (strong) measurement leads to a finite time Zeno effect trapping
the qubit's electron in one of the quantum dots. Mathematical explanation
linking 
the effect to the presence of singular terms in the residence time 
amplitude distribution should also apply to a wide range of similar measurements.
\newline
DS is grateful to Max-Planck Institute for Physics
of Complex Systems (Dresden) for hospitality and financial support
and to Shmuel Gurvitz for many a stimulating discussion.   


\begin{thebibliography}{8.}
\bibitem{FOOT1} 
Consider a von Neumann
measurement of an operator $\hat{A}=\int|a\ra a\la a|da$ 
 by a pointer with position $y$ and the coupling $\hat{H}_{int} = -i\partial_y\hat{A}\delta(t)$. With the initial product state 
 $|i\ra\int G(y)|y\ra dy$
the probability  to find a pointer reading $y$ and the system in a state $|f\ra$ is given by 
$P^{f\leftarrow i}(y)=|\int  G(y-a) \la f|a\ra)\la a|i\ra da|^2$. 
Comparing this with Eq.(\ref{0.10})  shows that the number of tunnelled atoms $n$ plays the role of a (discrete) pointer position.
\bibitem{FOOT2} While the classical uncertainty $\Delta \tau$ reflects one's lack of knowledge of the actual value of $\tau$, $\Delta^q \tau$ arises from the interference between the paths, which leaves the 
the 'actual' value of $\tau$ indeterminate.
\end{thebibliography}
\end{document}